# A WEARABLE HAPTIC GAME CONTROLLER

Jacques Foottit, Dave Brown, Stefan Marks and Andy M. Connor

Auckland University of Technology, Auckland, New Zealand
wgk9328@aut.ac.nz
wpv9142@aut.ac.nz
stefan.marks@aut.ac.nz
andrew.connor@aut.ac.nz

## ABSTRACT

*This paper outlines the development of a wearable game controller incorporating vibrotacticle haptic feedback that provides a low cost, versatile and intuitive interface for controlling digital games. The device differs from many traditional haptic feedback implementation in that it combines vibrotactile based haptic feedback with gesture based input, thus becoming a two way conduit between the user and the virtual environment. The device is intended to challenge what is considered an "interface" and draws on work in the area of Actor-Network theory to purposefully blur the boundary between man and machine. This allows for a more immersive experience, so rather than making the user feel like they are controlling an aircraft the intuitive interface allows the user to become the aircraft that is controlled by the movements of the user's hand. This device invites playful action and thrill. It bridges new territory on portable and low cost solutions for haptic controllers in a gaming context.*

## 1. INTRODUCTION

There are two significant developments that have arisen which have done a great deal to influence how people interact with machines. One of these developments comes from the area of sociology, where the Actor-Network Theory [1] has been challenging the divide between the social and the technical. The other significant development is the rise of wearable technology to the point where it has left the realm of high-tech labs and made inroads into classrooms, fashion, arts, and hobbyist garages. These developments both point towards a fundamental shift in terms of technology becoming part of the human perception of self. For example, the prolific use of mobile phones has developed to the point where the use of such devices can be thought of in terms of prosthetic augmentation of the human user. Interaction with these devices has become so natural and fundamental to modern life that they are no longer just a part of the social interaction, but an invisible mediator to the communication between people.

The view of objects like mobile phones as an extension of a human is something that is increasingly relevant as other advances continue to blur the line between human and machine. Advances in prosthetics and orthotics have raised questions around the rights of the people who use them [2], and it is quite telling that the term 'cyborg' is used to describe such people. While prosthetics are defined as artificial artefacts that replace the normal functionality of a defective organ, orthotics in this emerging discussion refers to artificial artefacts that add some capability to the human user. With this definition in mind, a wearable haptic feedback device can be viewed as an orthotic that adds the capability of interaction with machines that is not present naturally. The capability it adds is the ability to communicate with machines using gesture and to sense the virtual world with tactile sensations. This is particularly useful considering the reliance humans have on touch and gesture when communicating with each other.

Before children can speak, they explore the world through touch. Research into childhood development has shown that gestures are not only learned before language, but even support and influence the development of language [3]. Although the entire body as a whole communicates a great deal, the limb of choice for expression and exploration through gesture and touch is the hand. The anatomy of the hand makes it uniquely suited to such activities. The versatility and range of sensation available means the hand can interact with and experience the world better than any other part of the body. Hand gestures can even become a language in their own right, as in the case of sign language. Given the natural and fundamental nature of this mode of interaction, it makes sense to utilise the hands for interaction with machines. However while many machine interfaces utilise the hands for interaction, there is often a very clear divide between the hand and the machine interface. This places the interface as a physical intermediary between the human and the machine, requiring the human to translate their intentions into the interface language. In order to create a more intuitive interface, it becomes necessary to shift some of the translation work from the human to the machine.

The goal of this research is to explore a way of interacting with machines that minimises the layer of translation by focusing directly on the human hand as an input/output device. This is achieved through the use of a haptic feedback glove device that is designed to become an orthotic for the user. By minimising the perception of an intermediary interface, the haptic feedback glove enables the user to more naturally and intuitively communicate with a computer by acting as a tangible interface [4]. In particular, this development of the glove is focused in the first instance as an intuitive and useable game interface. Other low cost systems have been developed [5] that combine gestures and haptic feedback which indicate that users preferred vibrotactile feedback when interacting with virtual environments. However, such work only describes the outcomes of the work and not the development process. This paper focuses on the documentation of the development of such an interface in the context of a wider study.

This paper is an extended version of previously published work[1] and is structured as follows: In section 2, we discuss the motivation of the work and provide an overview of the related work. Section 3 presents the outcomes of a systematic literature review used to identify wearable haptic devices currently being developed and confirm there is a niche for a new device. In Section 4 we outline the development processes used and describe the final prototype glove and in Section 5 we describe how the glove is deployed as a game controller. Section 6 outlines possible future directions and Section 7 concludes the paper.

## 2. BACKGROUND AND RELATED WORK

The Actor-Network-Theory as proposed by Latour [1] is a relatively new theory in the field of sociology. The core of this new theory is the attempt to combine the idea of technological networks and social networks into a unified system. This arises from the observation that the technical and the social cannot truly be viewed in isolation. As a result, technological objects become part of the same network as people, along with virtually all other objects. This leads to the Actor-Network-Theory broadening the view of 'social' to include all associations. One outcome of this is that it becomes a kind of recursive system where an actor in a network is itself a network of actors. In addition these networks are dynamic and emergent rather than material, so it is possible for actors to dynamically merge and split to form new networks.

---

[1] J. Foottit, D. Brown, S. Marks, and A. M. Connor, "An Intuitive Tangible Game Controller," presented at The 10th Australasian Conference on Interactive Entertainment, Newcastle, Australia, 2014.

Latour's claims in Actor-Network Theory can be somewhat incredible upon first glance, however it is impossible to deny the very real connections between humans and the technology they create. Technology is created to support and interact with humans, and at the same time new technologies are shaping the way that humans interact with each other. Latour is not alone in his view of technology either. Amber Case presents a similar view when she says "The most successful technology gets out of the way and helps us live our lives. And really, it ends up being more human than technology, because we're co-creating each other all the time." [6]. Features like Siri, Apple's virtual assistant, epitomise this underlying goal of creating more human technology. Here the interaction is intentionally designed to be more than just conveying information. The virtual interface is given a personality, and speaks with language that we associate with other humans more than with computers. At the same time, technology is becoming even more integrated into the human experience. Technologies like the mobile phone have not only radically changed the way people communicate but increasingly are becoming more like augmentations, blurring the line between the human actor and the machine. This creates a kind of dichotomy where on the one hand the device becomes like an invisible extension of a person, but on the other hand it becomes more human itself as when the person interacts with it to find the way to a good restaurant. This is a key concept when considering orthotics or prostheses as it helps explain how they are not an external object but an extension of the person.

As technology takes on more human features, the computer itself can easily be seen as a social agent in human interactions. While simpler objects such as toasters still have associations as part of the network of associations posited by Actor-Network Theory, they lack the depth of interaction that is experienced between people. In contrast, modern computers are increasingly able to produce complex interactions that more closely resemble the type of interactions that occur between people. As computers continue to develop and provide more complex and human-like interactions, the claim of Actor-Network Theory that these devices are themselves actors in the network of human social interaction may even become a truism. It makes sense then that the methods of interacting with computers begin to more closely resemble human to human interactions.

When it comes to human-machine interaction, the concept of an orthotic as defined by Clarke [2] opens up many possibilities. Clarke defines an orthotic as "an artefact that supplements or extends a human's capabilities". Using this definition, if providing input to a computer is considered an extension of the natural abilities of a human then computer input devices can be thought of as orthotics. In such a case the orthotic provides the ability for interaction with a previously inaccessible world, much the same way that an artificial eye might allow the user to see infrared radiation. Perhaps the more significant event is not the use of the device as an orthotic, but the change in perception when that orthotic becomes perceived as a part of the human user. This significant shift in perception does not seem to occur immediately, but rather happens when the user becomes so familiar with the artefact that it feels like a part of their body like an extra limb. Neil Harbisson [7] speaks of this key transition in perception when he describes his experience of hearing colour through an artificial interface that translates colour into audible tones.

In order to achieve the shift in perception of an artefact from an external tool to an extension of our own body, it is key that it somehow integrates into our body. This helps explain the recent rise of interest in wearable technology as devices became small enough to make such a thing possible. Simply having a device attached to the body allows it to become integrated into the user perception significantly more easily than a separate fixed device. Even hand held devices tend to be viewed as external tools more than their wearable alternatives. Wearable devices also have the advantage of allowing the user to interact with them for long periods of time without the need to consciously be aware of them. This allows the user to become accustomed to the presence of the device very quickly, perhaps even to the point where it seems uncomfortable or unnatural to be without it. This is the point at which the user is beginning to see the device as an extension of themselves.

## 2.1. Haptic Technology

Haptic technology has found acceptance across a broad spectrum of fields, from highly sophisticated simulators for training surgeons to everyday devices like mobile phones [8]. The diversity of applications also means there is a great diversity in the forms of implementation. Haptic feedback technology can be broadly divided into two main categories.

The first category consists of literal real-world simulations where the aim is to recreate the experience of touching real-world objects utilising an artificial interface [9]. This could include some adjustment of the real-world experience such as amplifying extremely small forces to make them perceptible to a human operator. Whilst many authors have argued that desktop haptic feedback is of little practical significance, the development of many wearable devices in this first category are still bulky and obtrusive [10]. The second category for haptic feedback technology is more of an abstract simulation of the real world. In this case the haptic feedback provides information to the user that is not a direct representation of the forces seen in the real world. For example haptic feedback can be used to convey emotion [11], or to draw the attention of a user [12].

These two broad categories each have a range of implementations associated with them, however the first tends to rely on force reflecting interfaces that are capable of applying constant position-based forces on the user. Such implementations are relatively rigid, unwieldy and heavily steeped in mathematical rigour. In contrast, implementations in the second category vary considerably in terms of their size, scope and means for providing haptic feedback. When not attempting to simulate a real world event, force reflecting interfaces are rare. Force reflecting interfaces are capable of providing highly realistic representation of real-world experiences, as they apply constant forces to the user based on calculated forces from the simulation. However this fidelity comes at a cost, and often these interfaces are large, heavy and power consuming. While sometimes force-reflecting interfaces are augmented with specific vibration or skin stretch actuators, they can also be used to provide vibration directly since vibration is simply a special case where the forces applied change rapidly. These interfaces are utilised for high end applications where real world simulations are the objective, such as virtual simulators for training surgeons [13]. The high fidelity required by these applications justifies the high cost, weight and size of such interfaces. However there remains a wide range of opportunities for haptic technology to be applied where the limitations of force reflecting interfaces make their implementation infeasible. This is where the second category of haptic feedback technology emerges.

For many applications of haptic feedback true, high fidelity representations of contact forces are not possible due to the cost, weight, and size involved. This is particularly true where the haptic interface needs to be portable and wearable.

## 2.2. The Importance of Gesture

Humans naturally use multimodal communication to great benefit when cooperating to achieve a common task [14, 15]. The modes of communication used involve multiple forms of signals including auditory, visual and haptic. These modes of communication are selected and combined to allow for clearer communication in a variety of situations with varying environmental conditions. For example, in a noisy environment the haptic or visual modes of communication can be used predominantly to compensate for the inability for auditory communication to carry the desired signal reliably. Multiple modes of communication also allow the receiver to compare the signals being received from the different modes to verify the integrity of the overall signal. This has many benefits, particularly when it comes to avoiding conflict and mistakes.

Developmentally, the haptic language is the first to appear, and has even been shown to support and influence the development of spoken language [3]. This tendency to use haptic language as a structure for learning spoken language continues into adult life. For example, when learning a new language a person may point to an object they wish to learn the word for the same way that a child may point at an object and have their parents respond by telling them the name of the object. The intuitive nature of the haptic language makes it ideal for these supportive purposes not only in communication between humans, but also in human-machine interaction. The benefits of gesture based communication with computers becomes even more evident when the computer is considered a social agent similar to another human.

### 2.2.1. Gesture in Virtual Presence

Online environments like massively-multiplayer online role-playing games (MMORPGs) have been identified as a potential research platform for social sciences and clinical therapy [16]. This potential is reliant on the interactions between people virtually having the same properties as interactions in physical spaces. This appears to be the case, with Yee et al. showing that established findings of interpersonal distance and eye gaze transfer reliably into the virtual environment. This is despite the rather different modality of movement employed by these environments at this time. The addition of modalities that more closely resemble physical interactions allows for even more fidelity between virtual and physical interactions.

Gesture based input also allows for more intuitive input where the virtual presence of the user is not human or even humanoid. For example, in a flight simulator the user still has a virtual presence, but in this case the user is represented by the aircraft. Gesture based input allows for a more immersive experience by allowing the user to become the aircraft rather than feel like they are controlling an aircraft. In the mind of a child, the hand can become any range of objects, and by taking gesture based input it is possible for the hand to virtually become anything again. In our installation, we demonstrate this by implementing a basic flight simulator that is controlled by the movements of the user's hand. The movements are designed to feel both natural and intuitive in order to promote a sense of immersion that would be difficult to realise with an alternative interface. In this example the user's hand can become the aircraft much the same way that a child would imagine it.

### 2.3. Hand-Based Interfaces

There has been a great deal of exploration regarding alternative interfaces for human machine interaction, particularly with the improvement of sensing devices and processing power of computer systems. Hand based interfaces are not entirely new, with examples dating back to the 1980s. For example, Zimmerman et al [17] describes the development of a hand gesture interface device that utilises multiple sensors to track hand position and gestures. These early interfaces are often heavily reliant on inverse kinematic models embedded in software and often require extensive calibration before use. Whilst the interface may ultimately be quite effective, the need for calibration reduces the intuitive nature and emphasises the fact that the device is an external object rather than an invisible intermediary.

In many cases the hands have been focused on as a primary means of interaction without the need for a specific device, for example the SmartSkin [18] system uses interactive surfaces that are sensitive to human hand and finger gestures. As with the early devices based approaches, there are limitations to these gesture approaches in that they lack the ability to provide any form of haptic feedback to the user and therefore do not necessarily provide any higher degree of engagement than traditional interfaces. More recent approaches have looked at directly stimulating the skin [19] with relatively small devices such as robotic wearable thimbles. Whilst this approach has been successful, the thimbles are not yet sufficiently small to be feasible to stimulate all fingers on a hand without becoming obtrusive. As a result, such systems that utilise multiple thimbles [20] do not necessarily break down the barriers between man and machine.

Attempts have been made to combine the advantages of device based approaches with the advantages of gesture based approaches. For example, the Charade system [21] utilises a tethered glove that interprets finger positions and hand orientations in the context of a heavily scripted gesture language. Again, there is no implementation of tactile feedback to the user to indicate what events have been successfully interpreted. This limitation has been addressed by other work [5] that combines gestures and haptic feedback using fingertip devices in experiments comparing the user experience in a virtual environment. The outcomes of these experiments indicated that users preferred the vibrotactile feedback. Whilst the authors of this work indicate that their approach is low cost, at least in comparison to the CyberGrasp system, there is little explanation of how to produce a low cost system.

In contrast, other approaches have been developed that do provide such feedback but lack the ability to interact with a digital environment. For example, Frati & Prattichizzo [22] explore the suitability of combining gesture tracking using a Kinect with haptic feedback, though do so by having a literal hand in the virtual environment. To date, there has been little work conducted that is based on using the real hand to control objects that are not part of the human anatomy. Such approaches have been shown to provide different concepts of engagement and play [23]. This is just one of many systems that utilise some form of camera for tracking, with many others discussed in the literature [24-26].

A very limited number of attempts have been made to integrate gesture tracking into game controllers, particularly with haptic feedback. Tindale, Cumming, Pridham, Peter & Diamond [27] outline the development of a number of wearable haptic game devices, though these are self-contained gaming environments with a predefined mode of interaction. In contrast, Ionescu et al [28] describe an attempt which involves gesture tracking for one hand whilst using a game controller in the other. Such approaches seem unwieldy and further emphasise that the controller is an external object, not an invisible intermediary between man and machine.

Outside of academic research, a number of commercial systems are in production. A novel approach to user interfaces focusing on the hand is an interface called Thumbles[2]. This takes the approach of using physical objects to represent virtual controls. The novel part of this approach is that physical artefacts change to suit the virtual environment. They move around and can be interacted with by the user to alter all kinds of controls. This emphasises the importance of the human hand, but also focuses on the physical nature of an interface in preference to immaterial virtual interfaces. A haptic feedback glove has the potential to add a level of physicality to a virtual interface without needing to resort to physical objects that can become distracting and limited. Rather than altering the physical interface, the aim of haptic gloves is to make the physical interface invisible to the user, immersing them into an intuitive virtual environment that can be as dynamic and varied as the imagination allows.

The Myo[3] is another example of a modern approach to human machine interaction. This device also focuses on the hands as a primary means of interaction. In this case, the device measures electrical signals to the muscles of the hand to allow for gesture based input. It is also capable of providing haptic feedback, however the location of the device on the arm removes the haptic feedback from where it is most relevant – the fingertips. While the low profile and light weight of the device makes it ideal for freedom of motion, a glove can offer these same advantages as well as providing a platform for haptic feedback at the point where people are accustomed to receiving it. A glove is also something that people are very familiar with, and so long as it does not impede motion of the hand it can quickly be forgotten that it is even being worn.

---

[2] http://www.pattenstudio.com/projects/thumbles/
[3] https://www.thalmic.com/en/myo/

The range of interfaces emerging provides a range of options for users, each with their own benefits and limitations. In most cases, existing hand based interfaces either provide some form of haptic feedback or they provide some means of gesture tracking or user input. To our knowledge, it appears that there has been no attempt to provide both haptic feedback and gesture tracking in the same device, particularly when that device is intended to be an "invisible interface".

Different situations are likely to require different solutions, however the versatility of these emerging technologies means that more intuitive interaction with machines is becoming accessible in an ever increasing range of environments. The goal of this project is to develop a device that operates in this manner and provides a successful integration of different technologies. The development of the wearable haptic glove outlined in this paper has been further informed through the conduct of a systematic literature review to identify emergent devices. This review is outlined in the following sections.

## 3. SYSTEMATIC REVIEW OF DEVICES

This paper aims to survey the latest developments in portable wearable haptic interfaces in order to identify the current capabilities that can be achieved with such devices and the applications that are possible. For the purposes of this paper, the term "wearable" refers to a device where the entire device interfaces with the human body, including devices that are entirely hand held. This definition is chosen to include devices such as game controllers that are designed to interface with the human hand in a way that they become a part of the person rather than an externally mounted device.

### 3.1. Review Method

This review was conducted by searching IEEE Xplore, ASME, SpringerLink, Wiley, CiteSeerX, PubMed and ACM databases using the search terms 'Portable Wearable Haptic' and 'Portable Haptic Glove'. Results were limited to research published in the years 2012 – 2014 to focus only on developments in this field relevant to the conduct of this study which was undertaken in 2014. We acknowledge the limitations of this approach as haptic devices have been under development for a much longer timescale. To compensate this, the systematic review has been augmented by a more general review of devices to identify similar approaches in the broader literature which have already been discussed in Section 2.3.

Each result was then assessed for suitability based on inclusion and exclusion criteria, resulting in a final list of 16 primary sources. The inclusion criteria required the source paper to present a device that could be entirely interfaced with the human body, provided haptic feedback, and could be untethered. The inclusion criteria did not require the device to be presented in an untethered form, it only required the possibility of untethered operation. This allowed for connection of the device to a computer for the purposes of data gathering often associated with such research while still excluding devices that utilise externally mounted drive systems such as high power motor drivers and pneumatic systems that cannot be interfaced with the human body.

The exclusion criteria filtered out sources that were more than 2 years old, that presented non-wearable devices, and papers that did not present a complete haptic device. The final exclusion criteria disregarded papers that presented only an actuator or sensor without it being mounted and used in a haptic application. It did not exclude limited prototypes that focused on a new technology such as implementing novel actuation of a single finger instead of a full hand. As noted previously, for the purposes of this review the term "wearable" refers to a device where the entire device can be interfaced with the human body. This requires all the components of the haptic device to be able to be held or worn by the user, but does not require untethered operation and allows for hand held devices such as game controllers. The control system was also not required to be wearable, only the haptic device. This allowed for a regular desktop computer to provide the simulated environment. In cases where a wearable device was attached to a non-wearable device such as a glove being attached to desktop mounted force reflecting manipulandum, the paper was included but only the wearable device was considered.

### 3.2. Reviewed Devices

Of the sixteen primary sources included in the review, ten implemented vibration feedback without force feedback, five implemented force feedback without vibration, and one implemented both vibration and force feedback. Five devices implemented custom designed actuators. Eight of the devices aimed to simulate real forces acting upon the human body, while eight utilised haptic feedback for abstracted communication of information. Nine devices provided haptic feedback to the hand, three provided feedback to the arm and wrist, three provided feedback to the torso and arms, and one provided feedback to the head. Nine devices were created for specific domains and seven were for general or unspecified domains. The specific application domains included aged care [29, 30], health care [31], arts [32, 33], psychology [34, 35] and accessibility [36, 37]. 1 of the included devices was completely passive, with a tactile button placed at a fixed position in relation to the hand providing feedback [38].

#### 3.2.1. Wearable Force Feedback Devices

The seven devices that implemented some form of force feedback did so in varying ways and with different levels of fidelity. Two of the devices provided force feedback to the fingertip by utilising a custom design where a small platform in contact with the finger was acted on with either three geared motors [39] or three microservos [40]. This allowed force to be applied to the finger in different directions to give three degrees of freedom. Another device utilised a small belt attached to two geared motors to apply skin stretch due to friction [41], as well as applying vibration and contact force through a voice coil motor. While these devices presented promising results, they were generally quite bulky and only able to be implemented on one or two fingers due to their size. There was also a full exoskeleton based glove [42] that could provide force feedback to the fingers with one degree of freedom utilising a cable drive actuated by DC motors and worm gears. This system was capable of providing 10N of active force and was also mechanically locked due to the nature of the worm gears, providing more than 35N of passive force. Force feedback was also employed by one device based on a surgical implement that is hand held [31]. In this case, the force applied at the tip of the implement is amplified and fed back to the operator at the hand utilising a voice coil motor. This provided a one degree of freedom device specifically designed for surgical applications. All the devices that provided force feedback utilised custom designed actuators.

#### 3.2.2. Wearable Vibrotactile Feedback Devices

Vibrotactile feedback was utilised for eleven of the devices included in the review. Six of these devices utilised vibrotactile feedback exclusively, while one implemented vibration as part of the force actuation [41] and four implemented other forms of feedback in addition to vibration. Other forms of haptic feedback included heat [33], pressure [29], and electrical muscle stimulation [43]. Vibrotactile feedback was used in a wide variety of applications compared to force feedback, and also was mostly used to portray abstracted information rather than literal simulations of touch. For example, one of the devices was designed to vibrate when a child with ADHD lost focus on a task and thus restore their attention [34]. Two were designed to portray visual information such as distances from objects to a visually impaired user. One device utilised vibration to portray distances from objects [36], while the other utilised vibration to alert the user to obstacles [37]. The devices utilising vibration also varied in position on the body, including the torso [32, 33, 36], arms [29, 34, 43] and head [37]. In contrast, all the devices utilising force feedback were mounted on the hand. The majority of devices utilised off the shelf components, which included a vibrotactile glove that was exclusively comprised of components from off the shelf [44]. However, one of the devices incorporated a custom designed actuator to produce vibration [35]..

Whilst some of the devices reviewed utilise low cost components, most are too expensive to be considered as potential game controllers. Only one device [44] is sufficiently low cost and provides enough opportunity to be a useful inspiration. However, this device has a number of limitations. Whilst the device operates as a bi-directional interface, it is considerably bulky and obtrusive. The inventors of the device in questionargue that it "provides a good solution to the challenge of designing wearable haptics, firstly that it is to be small and light, secondly that it provides adequate dexterity without constraining hand motions, and thirdly that is has sufficient dynamic range to be versatile enough to be used in both very sensitive activities and in large force situations" [44]. Whilst we acknowledge this contribution, a game controller does not necessarily need to deal with large forces and therefore removing the need to address the challenge of dynamic range suggests that there is a potential to develop a smaller and lighter, low cost, wearable haptic game controller. In particular by aiming to also reduce the obtrusiveness, this would be the first step towards more immersive and tangible interfaces to a variety of digital environments.

## 4. A WEARABLE HAPTIC GLOVE

In this section, multiple facets of the development of the glove are outlined with a focus on the development methodology, informal feedback from the development process regarding the human experience and also outlining the glove itself.

### 4.1. Development Methodology

One of the core design principles utilised in this project was the use of rapid prototyping to quickly iterate through possible designs. The first weeks of the project were spent testing a variety of sensors, actuators and physical designs to develop a broad understanding of the range of options available. This process proved to be very effective at allowing us to quickly rule out technologies that would be ineffective, expensive, or bulky. It also revealed limitations to the technologies that showed promise, which in turn influenced the goals of the project. For example, initially we set out to create a haptic feedback glove that could provide position based force reflection but after our initial research into existing systems and technologies available decided to aim for a portable, lightweight system at the expense of fidelity.

The rapid prototyping philosophy worked so effectively that we continued to use it during our second phase of design where our first full prototype was refined and enhanced. This influenced the process of the project significantly, as we had originally planned that we would build the glove and then design an implementation to suit it. Instead we decided to develop the implementation and the glove in parallel, utilising rapid prototyping to allow the two aspects of the project to feed into each other.

### 4.2. Integrating Technology and Human Experience

Throughout the development of the glove we explored how to integrate technology into the human experience. It is a curious thing when a device becomes so natural that it is almost like an extension of the person. There are many advances in user interfaces that aim towards making the technology more natural and intuitive, but achieving a truly integrated experience where the interface becomes part of a person's experience of themselves remains rare. At the forefront of this integration of technology and humanity is prosthetics and orthotics, technological devices designed specifically to integrate with the human body. Devices like mobile phones and cars that radically transform the way a person can interact with the world around them also tend to become integrated into the human experience over time, although these devices are much less likely to become a part of the user's perception of themselves the way an orthotic or prosthetic might.

A common thread emerges when looking at the technologies that successfully integrate into the human experience – the technologies must fit well with the human body. In our project, it quickly became clear that it would be vital for our glove to fit the hand comfortably and be light enough not to impede the mobility of the hand. Wireless communication was also important for our project, as being tethered to a computer creates a physical and psychological barrier that separates the technological device from the user's perception of themselves. The difference between using the glove while wired to a computer and using it with the Bluetooth connection and battery power was remarkably distinct. A great deal of this change in the experience came from the need to keep track of the cable when using the wired solution. It distracted from the user experience by requiring the user to be aware of the position of the cable in order to avoid it getting tangled or pulling out. The lightness and small size of the glove was also a focus, and using a custom fabricated PCB and a small form factor Arduino contributed a great deal towards achieving our goals in this area. The use of a custom knitted glove was also significant, as it was far more comfortable than the earlier prototypes. This was primarily due to the flexible nature of the fabric that still held the optical sensors in place. Again this was an iterated process and the comfort of the glove was a prime consideration during development.

### 4.3. The Glove

The key to wearable technology is the miniaturisation of technology. As devices become smaller and more energy efficient, it becomes possible to embed them into worn artefacts. In the case of providing input to a computer, a particularly useful piece of technology that has developed greatly in recent years is the Micro-Electro-Mechanical Systems (MEMS) based Inertial Measurement Unit (IMU). These remarkable units allow for the combination of sensors such as accelerometers, gyroscopes and magnetometers into incredibly small form factors and were a key aspect of the final glove design. In fact, these devices can be as small as a few millimetres in length, width and height. They are also increasingly affordable, and have become ubiquitous in mobile technology such as smart phones. Although there has been much improvement in this area, these small, affordable devices are still considered somewhat inaccurate compared to their more expensive and bulky counterparts. However their accuracy is sufficient for most consumer applications, making them an ideal solution for providing orientation information for a glove-based input device. Their key limitation is a tendency to drift over time – particularly when being used to provide positional information. To overcome this limitation, technologies like the Microsoft Kinect that provide absolute position and orientation information can be used to supplement the data provided by these devices [45].

As has been mentioned in section 4.1, a rapid prototyping methodology was used in the development of the glove. Early experiments in the development process explored the performance of different sensors and actuators in isolation that quickly led to an understanding of the design space. Examples of experiments conducted in this stage of initial prototyping included the possible use of "muscle wire" as a means of providing a compressive force to the fingers (Figure 1), using 3D printers to test designs for mounting hardware on or near the fingertips (Figure 2) and designing low cost flex sensors.

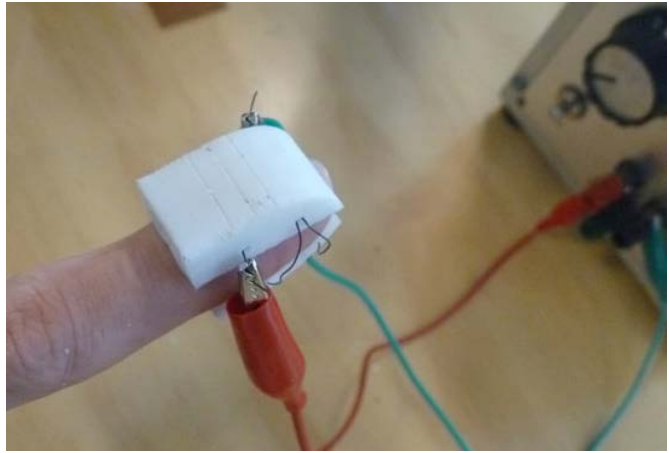

Figure 1. Muscle wire experiment

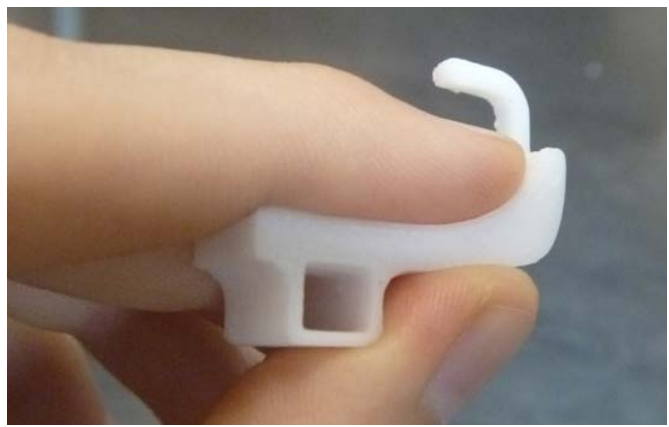

Figure 2. 3D printed fingertip hardware mount

The flex sensor is show in Figure 3 and utilises a piece of tubing with an infra-red LED at one end and an infra-red receiver at the other. Clear changes in resistance across the receiver were observed as the tube was bent.

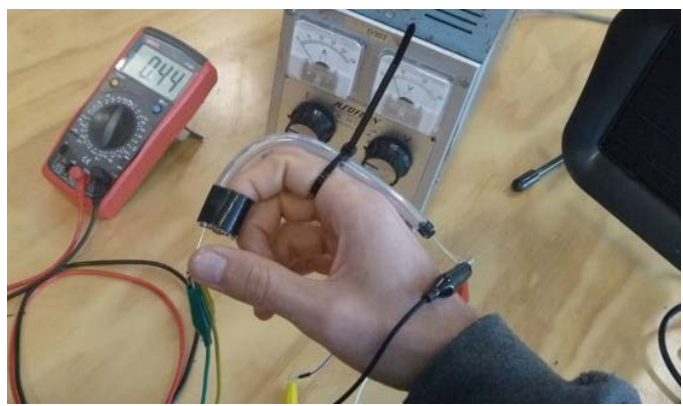

Figure 3. Testing the flex sensor

Throughout this testing phase, various partial prototypes were developed that provided insight into how best to integrate hardware into different glove "substrates" which informed the final design considerably. Such a partial prototype is shown in Figure 4 whilst testing the development of the connection through to the Unity game engine.

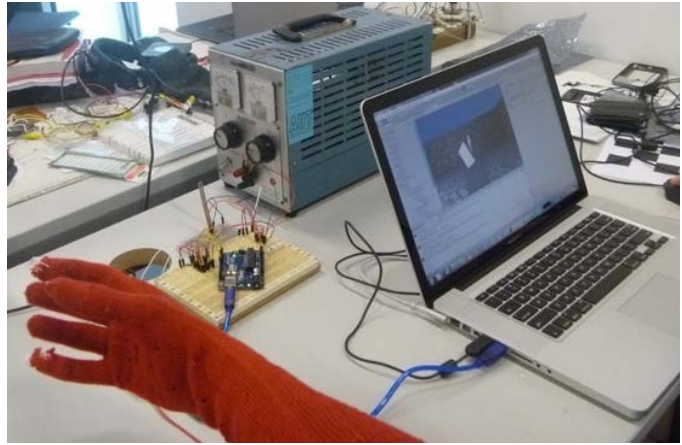

Figure 4. Partial prototype under test

Our initial prototypes focused on the development of input sensors, primarily due to the need to create a custom PCB for the motor drivers, initially circumvented by the use of a more bulky standard Arduino microcontroller and sandwich board based electronics. The first complete prototype is shown in Figure 5.

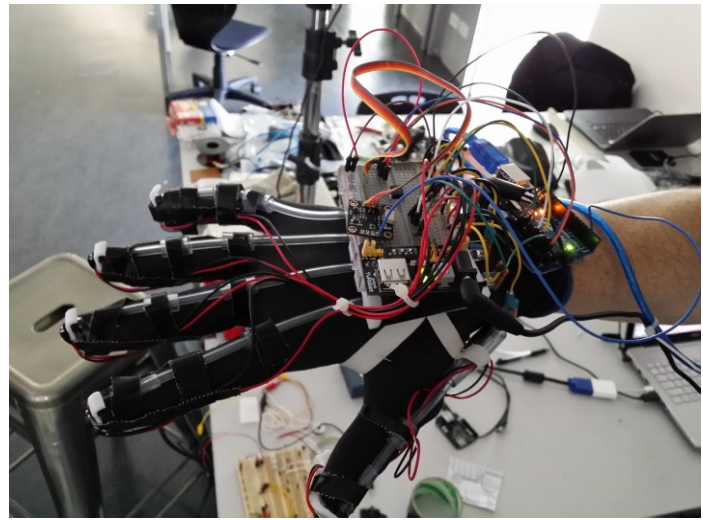

Figure 5. Initial completed glove prototype

Our final prototype includes a flex sensor for each of the fingers and an IMU feeding into Unity to control a virtual hand. An Arduino pro mini board is used to read the sensors and communicate with the computer. Vibratory motors are mounted in the fingertips of each of the fingers, and are triggered by events in the Unity program. In addition, the device can operate either by cable using a USB and power cable, or it can operate in a wireless mode by utilising a Bluetooth module and battery power. The final, complete version of the glove is shown in Figure 6.

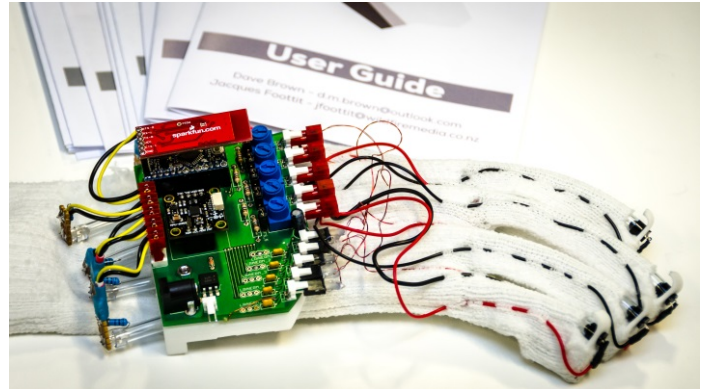

Figure 6.  Final custom knit glove and custom PCB

## 5. THE GAME ENVIRONMENT

The potential for the haptic glove to operate as an effective and intuitive interface has been demonstrated through the development of a basic flight simulator in Unity that is controlled by the movements of the user's hand. The required movements are designed to feel intuitive and allow for a sense of immersion that would be difficult to achieve with an alternative interface. The use of the hand as a controller for a game object is a key outcome of this work that is the result of the initial framing of the work in terms of the Actor-Network Theory. In this example the user's hand can become the aircraft in the same way that a child would imagine it. The flight simulator is controlled through a combination of the haptic glove and a Microsoft Kinect. The player moves their hand closer or further away from the Kinect to change the velocity of the aircraft with hand orientation changing aircraft placement. To fire guns the player has the choice of bending the thumb or clenching the fist to shoot missiles. The vibratory motors on the fingertips cause haptic responses to the player. The controls excel at generating a learning basis for finger movement and placement, intuitively forcing creative hand gestures and participation. Figure 7 illustrates the controls implemented by the flight simulator.

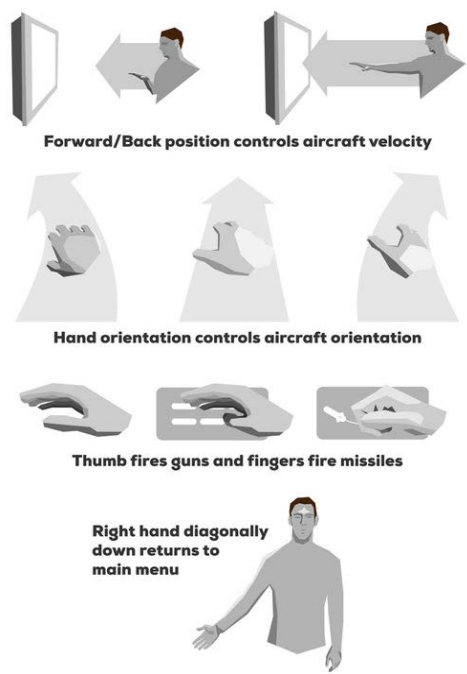

Figure 7.  Flight simulator controls

The game itself is rather limited: Whilst the player can control their own aircraft and attack enemy vessels, these vessels are essentially static entities that do not fight back. As it was implemented, the game also had no capacity for a player to crash into objects, the ground or to lose a life. Whilst perfectly functional as a demonstration of the haptic glove, the game would have little appeal as a game in its own right and would likely have limited ability to engage a player for more than a few minutes.

A formal usability study of the haptic glove has not yet been conducted. The glove is currently undergoing considerable refinement and formal evaluation will be postponed until this is completed. However, it is possible to draw anecdotal conclusions regarding the usage of the glove from the extensive play testing conducted during development and also from observations and discussions with users who trialled the glove as a game interface at a public demonstration. Figure 8 shows a player interacting with the game environment through the glove interface.

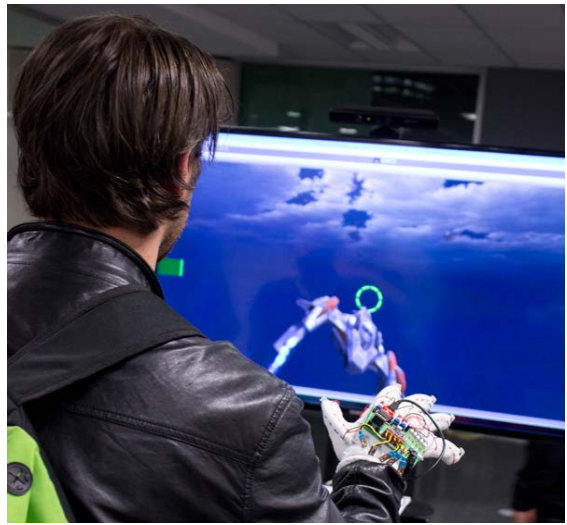

Figure 8. Gameplay

The glove was demonstrated at an exhibition of student work and was considered a huge success. Lots of people were interested in the project and had a chance to give the glove a go. The glove held up well with the frequent hand changing, and the flight simulator was very stable too. Most people quickly adapted to the use of the glove in the game without referring to the usage instructions, suggesting that the interface and controls are suitably intuitive. We got some great feedback on the night too - almost everyone commented that the haptic sensations were good, and we also got a few comments about the intuitive nature of the controls. In general, most players engaged with the game for much longer than would have been expected given the simple implementation. This, along with the feedback from users, suggests that the glove has the potential to be an intuitive and non-obtrusive game interface that supports increased engagement.

## 5. FUTURE WORK

Whilst the developed prototype has been a great success in terms of developing a two way tangible interface, there is still considerable scope to extend this work in new directions as well as focusing on refining the initial prototype. Vibrotactile feedback is an easily implemented form of haptic feedback readily utilised in portable wearable applications. While the force feedback devices remained focused on representing literal physical forces, a vibrotactile device uses more abstracted forms of input for the haptic interface. Vibrotactile haptic feedback can be very useful for conveying information to a user that is not intrinsically haptic in nature. It also indicates that for many situations true and realistic haptic simulations are not necessary. Where true and realistic haptic feedback is not required, vibrotactile feedback provides an easily implemented, affordable and versatile solution, however there is scope for further enhancement of the haptic glove concept.

A particular area of research that remains to be explored is the use of friction as a means of providing force feedback. This has the potential to be a low cost and unobtrusive means of providing a direct simulation of real world forces. One of the force feedback systems identified in the systematic review of devices utilised a cable-drive mechanism to provide force against the fingertips, however this system utilised worm gears to control the cable positions. This had the benefit of passively locking the fingers in place, but required the system to actively move with the fingers when not applying simulated force. In contrast, a system based on friction could allow the cables to freely move, and then apply friction to the cables to limit movement. This would remove the ability of the system to actively apply force to the fingers, but would have the benefit of only requiring power when friction needs to be applied. Whilst not strictly required for a game controller, the ability to provide force-feedback would allow the haptic glove concept to be extended to a broader range of virtual environments.

Another area for further research is the use of electrical muscle stimulation for providing haptic feedback. While there was one device included in the review that provided electrical muscle stimulation for haptic feedback [43], the results of the study indicated that further research was required. The lower power requirements for electrical muscle stimulation make it ideal for portable wearable devices, although there remain a number of challenges and health considerations that warrant further research.

Games are just one example of a virtual environment, however most games are still limited to desktop monitors where as other virtual environments have been extended into the physical space immediately surrounding the user. Whilst some researchers argue that gloves are unwieldy in terms of tracking gestures in virtual environments [46] this ignores the potential for the haptic feedback. Work is currently in progress that uses large scale motion capture technology in conjunction with an Oculus Rift to allow the interaction with large scale data visualisation [47]. One specific example is the visualisation of the neurons in a spatio-temporal artificial neural network, where a motion capture "mouse" can be used to select individual neurons. This experience can easily be augmented through the application of a haptic glove as can many others. Figure 9 shows the glove being used in an environment where the user can interact with a number of virtual objects.

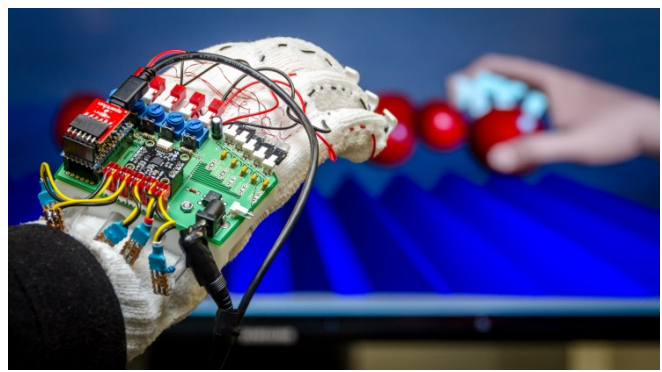

Figure 8. Object interaction simulator

The glove itself can also be extended to break the reliance on external motion tracking. Work is currently in progress to implement is a translation from wrist orientation to screen position so that the glove could act like a mouse input without the need for motion tracking.

However, the main direction for future work will entail a thorough and systematic usability evaluation of the glove. This will be undertaken when current revisions to the first prototype are complete and will entail not just investigating usability and engagement in the context of a use as a games controller but will also consider other types of digital and virtual environments.

## 5. CONCLUSION

The haptic feedback glove project explores alternative methods of interaction with machines by creating an artefact capable of providing gesture based input and haptic feedback. The potential applications of such technologies are widespread, and the benefits of this technology are still being discovered. The emphasis of the project was to create a device that integrates into the user experience to become an invisible mediator between the human and the machine. This enables a more intuitive and immersive experience for the user. To this end, the device is light weight and wireless at the expense of fidelity. The haptic feedback in particular is vibration based rather than force reflecting, allowing for low power and portability but making it impossible to restrict user motion based on virtual stimuli.

As this technology develops further, the interactions between humans and machines will more closely resemble the interactions between humans and each other. This has the potential to improve the immersion of virtual experiences, as well as reducing errors and improving input speed for a range of tasks. There is even the potential for this technology to allow for new methods of learning that are yet to be discovered.

**Authors**

Jacques Foottit obtained the Bachelor of Creative Technologies degree in 2014 from Auckland University of Technology in New Zealand. He is currently working towards his Master of Creative Technologies degree at Auckland University of Technology (AUT) in Auckland, New Zealand. His research interests relate to the use of haptic technology in virtual environments.

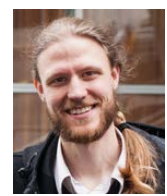

Dave Brown obtained the Bachelor of Creative Technologies degree in 2015 from Auckland University of Technology in New Zealand. His research interests include game development and interactive technology.

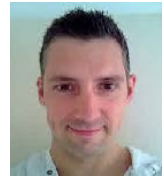

Dr. Stefan Marks is a researcher and lecturer at Colab, Auckland University of Technology. He holds a PhD from The University of Auckland for a project combining Virtual Environments, Serious Games and Image Processing into a medical teamwork simulation. He teaches in game programming, simulated and virtual environments. His research interests include virtual and augmented reality, serious games and computer graphics and visualisation.

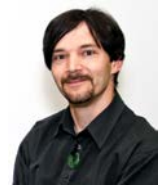

Dr Andy M. Connor obtained a Ph.D. in Mechatronics in 1996 from Liverpool John Moores University in the United Kingdom. He is currently employed as a Senior Lecturer in Colab, the interdisciplinary research/teaching nexus at Auckland University of Technology (AUT) in Auckland, New Zealand. His research interests span Software Engineering, Artificial Intelligence and Design Automation. Dr Andy M. Connor is a Member of the Institution of Engineering & Technology.

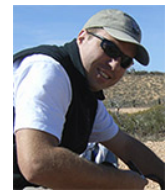